\documentclass[a4paper,oneside,reqno,12pt]{report}
\usepackage{amsmath,amssymb}
\usepackage{texdraw}
\usepackage{amscd,amsfonts}
\usepackage{amsthm}
\usepackage[ansinew]{inputenc}
\usepackage{amsbsy}
\usepackage[french]{babel}

\usepackage[dvips]{color}
\usepackage{graphics}
\usepackage{graphicx}

\newtheorem{thm}{Theorem}
\numberwithin{thm}{section}
\newtheorem{prop}[thm]{Proposition}
\newtheorem{lem}[thm]{ Lemma}
\newtheorem{cor}[thm]{ Corollary}

\theoremstyle{definition}

\newtheorem{nt}[thm]{Notation}

\newtheorem{dfn}{Definition}[section]

    \DeclareMathOperator{\cd}{\cd}

\newcommand{\dl}{\vskip.5cm}
\newcommand{\mrg}{\noindent}
\newcommand{\pr}{\dl\mrg\textit{ Proof. }}

\begin{document}

\noindent\textbf{\Large EXPRESSING A TENSOR PERMUTATION MATRIX $p^{\otimes n}$ IN TERMS OF THE GENERALIZED GELL-MANN MATRICES}\\

\noindent RAKOTONIRINA Christian\\
\noindent {\emph{Institut Supérieur de Technologie d'Antananarivo,
IST-T, BP 8122,\\ Madagascar}}\\
E-mail: rakotopierre@refer.mg

\begin{center}
\textbf{Abstract}
\end{center}
 We have shown how to express a tensor permutation matrix $p^{\otimes n}$
 as a linear combination of the tensor products
 of $p\times p$-Gell-Mann matrices. We have given the expression of
 a tensor permutation matrix $2\otimes2\otimes2$ as a linear
 combination of the tensor products of the Pauli matrices.

\section*{Introduction}
The expression of the tensor commutation matrix $2\otimes2$ as a
linear combination of the tensor products of the Pauli matrices
\begin{equation*}U_{2\otimes2}\;=\;\frac{1}{2}\;I_2\otimes I_2
+ \frac{1}{2}\sum_{i=1}^{3}\sigma_i\otimes\sigma_i\;
\end{equation*}
with
 $I_2$ the $2\times2$ unit matrix, \cite{fad95},  \cite{fuj01} are
frequently found in quantum theory. We had expressed the tensor
commutation matrix $p\otimes p$ as a linear combination of the
tensor products of the $p\times p$-Gell-Mann matrices \cite{rak05}.
In this paper, we show how to express a tensor permutation matrix
$p^{\otimes n}$ as a linear combination of the $p\times p$-Gell-Mann
matrices, with for $p=2$ the expression is in terms of the Pauli
matrices.
 \numberwithin{thm}{section}
\section{Tensor product of matrices}
\begin{dfn}
Consider $A=\left(A^i_j\right)\in\mathcal{M}_{m\times
n}\left(\mathbb{C}\right)$,
$B=\left(B^i_j\right)\in\mathcal{M}_{p\times
r}\left(\mathbb{C}\right)$. The matrix defined by
\begin{center}
$A\otimes B\;=\;\left(
\begin{array}{ccccc}
  A^1_1B & \ldots & A^1_jB & \ldots & A^1_nB \\
  \vdots &  & \vdots &  & \vdots \\
  A^i_1B & \ldots & A^i_jB & \ldots & A^i_nB \\
  \vdots &  & \vdots &  & \vdots \\
  A^m_1B & \ldots & A^m_jB & \ldots & A^m_nB \\
\end{array}%
\right)$
\end{center}
is called the tensor product of the matrix $A$ by the matrix $B$.
\begin{center}
$A\otimes B\in\mathcal{M}_{mp\times nr}\left(\mathbb{C}\right)$
\end{center}
\begin{center}
$A \otimes
B\;=\;\left(C_{j_1j_2}^{i_1i_2}\right)\;=\;\left(A^{i_1}_{j_1}B^{i_2}_{j_2}\right)$
\end{center}
(cf. for example \cite{fuj01} )
 where,\\
  $i_1i_2$ are row indices\\
  $j_1j_2$ are column indices.\\
\end{dfn}
\begin{prop}\label{thm11}
$\left(%
B_1\cdot A_1%
\right)$ $\otimes$ $\left(%
B_2\cdot A_2
\right)$\;=\;$\left(%
B_1 \otimes B_2
\right)\cdot\left(%
A_1 \otimes A_2
\right)$ for any matrices $B_1$, $A_1$, $B_2$, $A_2$ if the habitual
matricial products $B_1\cdot A_1$ and $B_2\cdot A_2$ are defined.
\end{prop}
\begin{prop}
Tensor product of matrices is associative.
\end{prop}
\begin{prop}
Tensor product of matrices is distributive with respect to the
addition of matrices.
\end{prop}
\begin{prop}
$I_n \otimes I_m$\;=\;$I_{nm}$
\end{prop}
\begin{thm}\label{thm12}
Consider $\left(%
 A_i%
\right)_{1\leq i\leq n\times m}$ a basis of
$\mathcal{M}_{n\times m}$($\mathbb{C}$), $\left(%
B_j %
\right)_{1\leq j\leq p\times r}$ a basis of
$\mathcal{M}_{p\times r}$($\mathbb{C}$). Then, $\left(%
 A_i\otimes B_j%
\right)_{1\leq i\leq n\times m, 1\leq j\leq p\times r}$ is a basis
of $\mathcal{M}_{np\times mr}$($\mathbb{C}$).
\end{thm}
\section{Tensor permutation matrices}
\begin{dfn}
For $p$, $q\in\mathbb{N}$, $p\geq 2$, $q\geq 2$, we call tensor
commutation matrix $p\otimes q$ the permutation matrix $U_{p\otimes
q}\in \mathcal{M}_{pq\times pq}\left(\mathbb{C}\right)$ formed by 0
and 1, verifying the property
\begin{equation*}
U_{p\otimes q}.(a\otimes b) \;=\; b\otimes a
\end{equation*}
for all $a\in\mathcal{M}_{p\times 1}\left(\mathbb{C}\right)$,
$b\in\mathcal{M}_{q\times 1}\left(\mathbb{C}\right)$.\\
 More generally, for $k\in\mathbb{N}$, $k>2$ and for $\sigma$ permutation
on $\{1, 2,\ldots, k\}$, we call $\sigma$-tensor permutation matrix
$n_{1}\otimes n_{2}\otimes\ldots\otimes n_{k}$ the permutation
matrix $U_{n_{1}\otimes n_{2}\otimes\ldots\otimes
n_{k}}(\sigma)\in\mathcal{M}_{n_{1}n_{2}\ldots n_{k}\times
n_{1}n_{2}\ldots n_{k}}\left(\mathbb{C}\right)$ formed by 0 and 1,
verifying the property
\begin{equation*}
U_{n_{1}\otimes n_{2}\otimes\ldots\otimes
n_{k}}(\sigma)\cdot\left(a_{1}\otimes a_{2}\otimes\ldots\otimes
a_{k}\right)=a_{\sigma(1)}\otimes a_{\sigma(2)}\otimes\ldots\otimes
a_{\sigma(k)}
\end{equation*}
for all $a_{i}\in\mathcal{M}_{n_{i}\times
1}\left(\mathbb{C}\right)$, $\left(i\in\{1, 2,\ldots, k\}\right)$.
\end{dfn}
\begin{dfn}
For $k\in\mathbb{N}$, $k>2$ and for $\sigma$ permutation on $\{1,
2,\ldots, k\}$, we call $\sigma$-tensor transposition  matrix
$n_{1}\otimes n_{2}\otimes\ldots\otimes n_{k}$ a $\sigma$-tensor
permutation matrix $n_{1}\otimes n_{2}\otimes\ldots\otimes n_{k}$
with $\sigma$ is a transposition.\\
\end{dfn}

Consider the $\sigma$-tensor transposition matrix $n_{1}\otimes
n_{2}\otimes \ldots\otimes n_{k}$, $U_{n_{1}\otimes n_{2}\otimes
\ldots\otimes n_{k}}(\sigma)$ with $\sigma$ the transposition
$(i\;\;j)$.
\begin{multline*}
U_{n_{1}\otimes n_{2}\otimes \ldots\otimes
n_{k}}(\sigma)\cdot\left(a_{1}\otimes\ldots\otimes a_{i}\otimes
a_{i+1}\otimes\ldots\otimes a_{j}\otimes a_{j+1}\otimes\ldots\otimes
a_{k}\right)\;
\\
=\;a_{1}\otimes\ldots\otimes a_{i-1}\otimes a_{j}\otimes
a_{i+1}\otimes\ldots\otimes a_{j-1}\otimes a_{i}\otimes
a_{j+1}\otimes\ldots\otimes a_{k}
\end{multline*}
for any
$a_{l}\in\mathcal{M}_{n_{l}\times1}\left(\mathbb{C}\right)$.\\
If $\left(B_{l_{i}}\right)_{1\leq l_{i}\leq n_{j}n_{i}}$,
$\left(B_{l_{j}}\right)_{1\leq l_{j}\leq n_{i}n_{j}}$ are
respectively bases of $\mathcal{M}_{n_{j}\times
n_{i}}\left(\mathbb{C}\right)$ and $\mathcal{M}_{n_{i}\times
n_{j}}\left(\mathbb{C}\right)$, then the tensor commutation matrix
$U_{n_{i}\otimes n_{j}}$ can be decomposed as a linear combination
of the basis $\left(B_{l_{i}}\otimes B_{l_{j}}\right)_{1\leq
l_{i}\leq n_{j}n_{i},1\leq l_{j}\leq n_{i}n_{j}}$ of
$\mathcal{M}_{n_{i}n_{j}\times n_{i}n_{j}}\left(\mathbb{C}\right)$.
 We want to prove that $U_{n_{1}\otimes n_{2}\otimes \ldots\otimes
n_{k}}(\sigma)$ is a linear combination of\\

$\left(I_{n_{1}n_{2}\ldots n_{i-1}}\otimes B_{l_{i}}\otimes
I_{n_{i+1}n_{i+2}\ldots n_{j-1}}\otimes B_{l_{j}}\otimes
I_{n_{j+1}n_{j+2}\ldots n_{k}}\right)_{1\leq l_{i}\leq
n_{j}n_{i},1\leq l_{j}\leq n_{i}n_{j}}$. \\
For doing so, it suffices us to prove the following theorem.

\begin{thm}\label{thm21}
Suppose \;$\sigma=\left(
                      \begin{array}{ccc}
                        1 & 2 & 3 \\
                        3 & 2 & 1 \\
                      \end{array}
                    \right)=\left(
                                  \begin{array}{cc}
                                    1 & 3 \\
                                  \end{array}
                                \right)$ permutation on $\{1, 2,
                                3\}$, $\left(B_{i_{1}}\right)_{1\leq i_{1}\leq
                                N_{3}N_{1}}$, $\left(B_{i_{3}}\right)_{1\leq i_{3}\leq
                                N_{1}N_{3}}$ bases respectively of $\mathcal{M}_{N_{3}\times
N_{1}}\left(\mathbb{C}\right)$ and $\mathcal{M}_{N_{1}\times
N_{3}}\left(\mathbb{C}\right)$. If \;$U_{N_{1}\otimes
N_{3}}=\displaystyle\sum_{i_{1}=1}^{N_{1}N_{3}}\displaystyle\sum_{i_{3}=1}^{N_{1}N_{3}}\alpha^{i_{1}i_{3}}B_{i_{1}}\otimes
B_{i_{3}}$, $\alpha^{i_{1}i_{3}}\in\mathbb{C}$, then\\
$U_{N_{1}\otimes N_{2}\otimes
N_{3}}\left(\sigma\right)=\displaystyle\sum_{i_{1}=1}^{N_{1}N_{3}}\displaystyle\sum_{i_{3}=1}^{N_{1}N_{3}}\alpha^{i_{1}i_{3}}B_{i_{1}}\otimes
I_{N_{2}}\otimes B_{i_{3}}$.
\end{thm}
\pr Let $b_{1}=\left(
                 \begin{array}{c}
                   \beta_{1}^{1} \\
                   \vdots \\
                   \beta_{1}^{N_{1}} \\
                 \end{array}
               \right)\in\mathcal{M}_{N_{1}\times
1}\left(\mathbb{C}\right)$, $b_{3}=\left(
                 \begin{array}{c}
                   \beta_{3}^{1} \\
                   \vdots \\
                   \beta_{3}^{N_{3}} \\
                 \end{array}
               \right)\in\mathcal{M}_{N_{3}\times
1}\left(\mathbb{C}\right)$.\\
At first let us develop the relation
\begin{equation*}
U_{N_{1}\otimes N_{3}}\cdot\left(b_{1}\otimes
b_{3}\right)=b_{3}\otimes b_{1}
\end{equation*}
Using the Proposition \ref{thm11} \numberwithin{equation}{section}
\begin{equation}\label{e21}
\displaystyle\sum_{i_{1}=1}^{N_{1}N_{3}}\displaystyle\sum_{i_{3}=1}^{N_{1}N_{3}}
\left(
  \begin{array}{c}
    \alpha^{i_{1}i_{3}}{B_{i_{1}}}_{j}^{1}\beta_{1}^{j}\left(B_{i_{3}}\cdot b_{3}\right) \\
    \vdots \\
    \alpha^{i_{1}i_{3}}{B_{i_{1}}}_{j}^{N_{1}}\beta_{1}^{j}\left(B_{i_{3}}\cdot b_{3}\right) \\
  \end{array}
\right)=\left(
          \begin{array}{c}
            \beta_{3}^{1}b_{1} \\
            \vdots \\
            \beta_{3}^{N_{3}}b_{1} \\
          \end{array}
        \right)
\end{equation}
where ${B_{i_{1}}}_{k}^{l}$ is the $l$-th row and $k$-th column of
the matrix $B_{i_{1}}$.\\
In the other hand, in using the Proposition \ref{thm11},
\begin{multline*}
\displaystyle\sum_{i_{1}=1}^{N_{1}N_{3}}\displaystyle\sum_{i_{3}=1}^{N_{1}N_{3}}\alpha^{i_{1}i_{3}}B_{i_{1}}\otimes
I_{N_{2}}\otimes B_{i_{3}}\cdot\left(b_{1}\otimes b_{2}\otimes
b_{3}\right)\\
=\displaystyle\sum_{i_{1}=1}^{N_{1}N_{3}}\displaystyle\sum_{i_{3}=1}^{N_{1}N_{3}}\alpha^{i_{1}i_{3}}
\left(B_{i_{1}}\cdot b_{1}\right)\otimes b_{2}\otimes
\left(B_{i_{3}}\cdot b_{3}\right)\\
=\displaystyle\sum_{i_{1}=1}^{N_{1}N_{3}}\displaystyle\sum_{i_{3}=1}^{N_{1}N_{3}}
\left(
  \begin{array}{c}
    \alpha^{i_{1}i_{3}}{B_{i_{1}}}_{j}^{1}\beta_{1}^{j}\beta_{2}^{1}\left(B_{i_{3}}\cdot b_{3}\right) \\
    \vdots\\
    \alpha^{i_{1}i_{3}}{B_{i_{1}}}_{j}^{1}\beta_{1}^{j}\beta_{2}^{l}\left(B_{i_{3}}\cdot b_{3}\right) \\
    \vdots \\
    \alpha^{i_{1}i_{3}}{B_{i_{1}}}_{j}^{1}\beta_{1}^{j}\beta_{2}^{N_{2}}\left(B_{i_{3}}\cdot b_{3}\right) \\
    \vdots \\
    \vdots \\
    \alpha^{i_{1}i_{3}}{B_{i_{1}}}_{j}^{N_{1}}\beta_{1}^{j}\beta_{2}^{1}\left(B_{i_{3}}\cdot b_{3}\right) \\
    \vdots \\
    \alpha^{i_{1}i_{3}}{B_{i_{1}}}_{j}^{N_{1}}\beta_{1}^{j}\beta_{2}^{l}\left(B_{i_{3}}\cdot b_{3}\right) \\
    \vdots \\
    \alpha^{i_{1}i_{3}}{B_{i_{1}}}_{j}^{N_{1}}\beta_{1}^{j}\beta_{2}^{N_{2}}\left(B_{i_{3}}\cdot b_{3}\right) \\
  \end{array}
\right)
\end{multline*}
\begin{equation*}
b_{3}\otimes b_{2}\otimes b_{1}=\left(
                                  \begin{array}{c}
                                    \beta_{3}^{1}\beta_{2}^{1}b_{1} \\
                                    \vdots \\
                                    \beta_{3}^{1}\beta_{2}^{l}b_{1} \\
                                    \vdots \\
                                    \beta_{3}^{1}\beta_{2}^{N_{2}}b_{1} \\
                                    \vdots \\
                                    \vdots \\
                                    \beta_{3}^{N_{3}}\beta_{2}^{1}b_{1} \\
                                    \vdots \\
                                    \beta_{3}^{N_{3}}\beta_{2}^{l}b_{1} \\
                                    \vdots \\
                                    \beta_{3}^{N_{3}}\beta_{2}^{N_{2}}b_{1} \\
                                  \end{array}
                                \right)
\end{equation*}
The matrices $\left(B_{i_{3}}\cdot b_{3}\right)$ are elements of
$\mathcal{M}_{N_{1}\times 1}\left(\mathbb{C}\right)$. Hence,
employing \eqref{e21} we have that if
\begin{equation*}
U_{N_{1}\otimes
N_{3}}=\displaystyle\sum_{i_{1}=1}^{N_{1}N_{3}}\displaystyle\sum_{i_{3}=1}^{N_{1}N_{3}}\alpha^{i_{1}i_{3}}B_{i_{1}}\otimes
B_{i_{3}},
\end{equation*}
then
\begin{equation*}
 U_{N_{1}\otimes N_{2}\otimes
N_{3}}\left(\sigma\right)=\displaystyle\sum_{i_{1}=1}^{N_{1}N_{3}}\displaystyle\sum_{i_{3}=1}^{N_{1}N_{3}}\alpha^{i_{1}i_{3}}B_{i_{1}}\otimes
I_{N_{2}}\otimes B_{i_{3}}.
\end{equation*}
 \qed
\section{Decomposition of a tensor permutation matrix}

\begin{nt}
    Let $\sigma\in S_{n}$, that is $\sigma$ a permutation on \{1, 2, \ldots, n\}, $p\in\mathbb{N}$, $p\geq2$, we denote the tensor
permutation matrix $U_{\underbrace{p\otimes p\otimes\ldots\otimes
p}_{n-times}}\left(\sigma\right)$ by $U_{p^{\otimes
n}}\left(\sigma\right)$,\cite{rao86}.
\end{nt}

We use the following lemma for demonstrating the theorem below
\cite{mer97}.
\begin{lem}Every permutation $\sigma\in S_{n}$ can be
written as a product of transpositions. (The factorization into a
product of transpositions may not be unique)
\begin{equation*}
\mathcal{C_{\sigma}}=\left(i_{1}\;i_{2}\;i_{3}\;\ldots
\;i_{n-1}\;i_{n}\right)=\left(i_{1}\;i_{2}\right)\left(i_{2}\;i_{3}\right)\ldots\left(i_{n-1}\;i_{n}\right)
\end{equation*}
\end{lem}
\begin{lem}\label{thm33}
Let $\sigma\in S_{n}$, whose cycle is
\begin{equation*}
\mathcal{C_{\sigma}}=\left(i_{1}\;i_{2}\;i_{3}\;\ldots
\;i_{n-1}\;i_{n}\right)
\end{equation*}
then,
\begin{equation*}
\mathcal{C_{\sigma}}=\left(i_{1}\;i_{2}\;i_{3}\;\ldots\;i_{n-1}\right)\left(i_{n-1}\;i_{n}\right)
\end{equation*}
\end{lem}
\begin{thm}\label{thm34}
For $n\in\mathbb{N}^{*}$, $n>1$, $\sigma\in S_{n}$ whose cycle is
\begin{equation*}
\mathcal{C_{\sigma}}=\left(i_{1}\;i_{2}\;i_{3}\;\ldots
\;i_{k-1}\;i_{k}\right)
\end{equation*}
with $k\in\mathbb{N}^{*}$, $k>2$. Then
\begin{equation*}
U_{p^{\otimes n}}\left(\sigma\right)=U_{p^{\otimes
n}}\left(\left(i_{1}\;i_{2}\;\ldots\;i_{k-1}\right)\right)\cdot
U_{p^{\otimes n}}\left(\left(i_{k-1}\;i_{k}\right)\right)
\end{equation*}
or
\begin{equation*}
U_{p^{\otimes n}}\left(\sigma\right)=U_{p^{\otimes
n}}\left(\left(i_{1}\;i_{2}\right)\right)\cdot U_{p^{\otimes
n}}\left(\left(i_{2}\;i_{3}\right)\right)\cdot\ldots \cdot
U_{p^{\otimes n}}\left(\left(i_{k-1}\;i_{k}\right)\right)
\end{equation*}
with $p\in\mathbb{N}^{*}$, $p>2$.
\end{thm}
So, a tensor permutation matrix can be expressed as a product of
tensor transposition matrices.
\begin{cor}
For $n\in\mathbb{N}^{*}$, $n>2$, $\sigma\in S_{n}$ whose cycle is
\begin{equation*}
\mathcal{C_{\sigma}}=\left(i_{1}\;i_{2}\;i_{3}\;\ldots
\;i_{n-1}\;n\right)
\end{equation*}
Then,
\begin{equation*}
U_{p^{\otimes n}}\left(\sigma\right)=\left[U_{p^{\otimes
(n-1)}}\left(\left(i_{1}\;i_{2}\;\ldots\;i_{n-2}\;i_{n-1}\right)\right)\otimes
I_{p}\right]\cdot U_{p^{\otimes
n}}\left(\left(i_{n-1}\;n\right)\right)
\end{equation*}
\end{cor}
\section{$n\times n$-Gell-Mann matrices}
The $n\times n$-Gell-Mann matrices are hermitian, traceless matrices
$\Lambda_{1}$,$\Lambda_{2}$,\ldots,$\Lambda_{n^{2}-1}$ satisfying
the commutation relations \cite{nar89}, \cite{pic95}
\begin{equation}\label{e41}
\left[\Lambda_{a},\Lambda_{b}\right]=2if_{abc}\Lambda_{c}
\end{equation}
where $f_{abc}$ are the structure constants which are reals and
totally antisymmetric, and
\begin{equation*}
Tr\left(\Lambda_{a},\Lambda_{b}\right)=2\delta_{ab}
\end{equation*}
where $\delta_{ab}$ the Kronecker symbol.\\
For $n=2$, the $2\times2$-Gell-Mann matrices are the usual Pauli
matrices, while for $n=3$, they correspond to the eight
$3\times3$-Gell-Mann matrices which are constructed by the following
way: the three first are the $3\times3$ matrices obtained in adding
to the three Pauli matrices third row and third column formed by
0, namely\\
 $\lambda_{1}=\left(
                         \begin{array}{ccc}
                           0 & 1 & 0 \\
                           1 & 0 & 0 \\
                           0 & 0 & 0 \\
                         \end{array}
                       \right)$
, $\lambda_{2}=\left(
                         \begin{array}{ccc}
                           0 & -i & 0 \\
                           i & 0 & 0 \\
                           0 & 0 & 0 \\
                         \end{array}
                       \right)$
, $\lambda_{3}=\left(
                         \begin{array}{ccc}
                           1 & 0 & 0 \\
                           0 & -1 & 0 \\
                           0 & 0 & 0 \\
                         \end{array}
                       \right)$\\
The next $3\times3$ matrices are obtained in adding to the two
non-diagonal Pauli matrices second column and second row formed by
0, namely\\
$\lambda_{4}=\left(
                         \begin{array}{ccc}
                           0 & 0 & 1 \\
                           0 & 0 & 0 \\
                           1 & 0 & 0 \\
                         \end{array}
                       \right)$,
$\lambda_{5}=\left(
                         \begin{array}{ccc}
                           0 & 0 & -i \\
                           0 & 0 & 0 \\
                           i & 0 & 0 \\
                         \end{array}
                       \right)$\\
The next $3\times3$ matrices are obtained in adding to the two
non-diagonal Pauli matrices first column and first row formed by 0,
namely\\
$\lambda_{6}=\left(
                         \begin{array}{ccc}
                           0 & 0 & 0 \\
                           0 & 0 & 1 \\
                           0 & 1 & 0 \\
                         \end{array}
                       \right)$,
$\lambda_{7}=\left(
                         \begin{array}{ccc}
                           0 & 0 & 0 \\
                           0 & 0 & -i \\
                           0 & i & 0 \\
                         \end{array}
                       \right)$\\
And finally, the last one is a diagonal matrix which is a
hermitian, traceless matrix with\\
\begin{equation*}
Tr\left({\lambda_{8}}^{2}\right)=2
\end{equation*}
namely\\
$\lambda_{8}=\frac{1}{\sqrt{3}}\left(
                                 \begin{array}{ccc}
                                   1 & 0 & 0 \\
                                   0 & 1 & 0 \\
                                   0 & 0 & -2 \\
                                 \end{array}
                               \right)\\
$ We can construct with the analogous way from the
$(n-1)\times(n-1)$-Gell-Mann matrices the $n\times n$-Gell-Mann
matrices. In doing so, the first $(n-1)^{2}-1$ $n\times n$-Gell-Mann
matrices are obtained in adding to each $(n-1)\times(n-1)$-Gell-Mann
matrices $n$-th column and $n$-th row formed by 0. The $(2n-2)$\;
$n\times n$-Gell-Mann matrices are the following non-diagonal
symmetric matrices and the following non-diagonal antisymmetric
matrices matrices\\
$\Lambda_{(n-1)^{2}}=\left(
  \begin{array}{cccccc}
    0 & 0 & \ldots & \ldots & 0 & 1 \\
    0 & 0 & &  &  & 0 \\
    \vdots &  & \ddots & &  & \vdots \\
    \vdots &  &  & \ddots &  & \vdots \\
    0 &  &  &  &  & 0 \\
    1 & 0 & \ldots & \ldots & 0 & 0 \\
  \end{array}
\right)$, $\Lambda_{(n-1)^{2}+1}=\left(
  \begin{array}{cccccc}
    0 & 0 & \ldots & \ldots & 0 & -i \\
    0 & 0 & &  &  & 0 \\
    \vdots &  & \ddots & &  & \vdots \\
    \vdots &  &  & \ddots &  & \vdots \\
    0 &  &  &  &  & 0 \\
    i & 0 & \ldots & \ldots & 0 & 0 \\
  \end{array}
\right)$, $\Lambda_{(n-1)^{2}+2}=\left(
  \begin{array}{cccccc}
    0 & 0 & \ldots & \ldots & 0 & 0 \\
    0 & 0 & &  &  & 1 \\
    \vdots &  & \ddots & &  & 0 \\
    0 &  &  & \ddots &  & \vdots \\
    1 &  &  &  &  & 0 \\
    0 & 0 & \ldots & \ldots & 0 & 0 \\
  \end{array}
\right)$, $\Lambda_{(n-1)^{2}+3}=\left(
  \begin{array}{cccccc}
    0 & 0 & \ldots & \ldots & 0 & 0 \\
    0 & 0 & &  &  & -i \\
    \vdots &  & \ddots & &  & 0 \\
    0 &  &  & \ddots &  & \vdots \\
    i &  &  &  &  & 0 \\
    0 & 0 & \ldots & \ldots & 0 & 0 \\
  \end{array}
\right)$,\\

 ...,\\

 $\Lambda_{n^{2}-3}=\left(
  \begin{array}{cccccc}
    0 & 0 & \ldots & \ldots & 0 & 0 \\
    \vdots & 0 & &  &  & \vdots \\
    \vdots &  & \ddots & &  & \vdots \\
    \vdots &  &  & \ddots &  & 0 \\
     &  &  &  & 0 & 1 \\
    0 & \ldots & \ldots & \ldots & 1 & 0 \\
  \end{array}
\right)$, $\Lambda_{n^{2}-2}=\left(
  \begin{array}{cccccc}
    0 & 0 & \ldots & \ldots & 0 & 0 \\
    \vdots & 0 & &  &  & \vdots \\
    \vdots &  & \ddots & &  & \vdots \\
    \vdots &  &  & \ddots &  & 0 \\
     &  &  &  & 0 & -i \\
    0 & \ldots & \ldots & \ldots & i & 0 \\
  \end{array}
\right)$\\
and finally, the last one is the diagonal matrix which is a
hermitian, traceless matrix with
\begin{equation*}
Tr\left({\Lambda_{n^{2}-1}}^{2}\right)=2
\end{equation*}
namely\\
$\Lambda_{n^{2}-1}=\frac{1}{\sqrt{C_{n}^{2}}}\left(
                                               \begin{array}{cccccc}
                                                 1 & 0 & \ldots & \ldots & \ldots & 0 \\
                                                 0 & 1 &  &  &  &  \\
                                                 \vdots &  & \ddots &  &  & \vdots \\
                                                 \vdots &  &  & \ddots &  & \vdots \\
                                                  &  &  &  & 1 & 0 \\
                                                 0 & \ldots & \ldots & \ldots & 0 & -(n-1) \\
                                               \end{array}
                                             \right)
$\\

They satisfy also the anticommutation relation \cite{nar89},
\cite{pic95}
\begin{equation}\label{e42}
\left\{\Lambda_{a},
\Lambda_{b}\right\}=\frac{4}{n}\delta_{ab}I_{n}+2\displaystyle\sum_{c=1}^{n^{2}-1}d_{abc}\Lambda_{c}
\end{equation}
where $I_{n}$ denotes the $n$-dimensional unit matrix and the
constants $d_{abc}$ are reals and totally symmetric in the three
indices, and by using the relations (\ref{e41}) and (\ref{e42}), we
have
\begin{equation}\label{e43}
\Lambda_{a}\Lambda_{b}=\frac{2}{n}\delta_{ab}+\displaystyle\sum_{c=1}^{n^{2}-1}d_{abc}\Lambda_{c}
+i\displaystyle\sum_{c=1}^{n^{2}-1}f_{abc}\Lambda_{c}
\end{equation}
The structure constants satisfy the relation \cite{nar89}
\begin{equation}\label{e44}
\displaystyle\sum_{e=1}^{n^{2}-1}f_{abe}f_{cde}=\frac{2}{n}\left(\delta_{ac}\delta_{bd}-\delta_{ad}\delta_{bc}\right)
+\displaystyle\sum_{e=1}^{n^{2}-1}d_{ace}d_{dbe}-\displaystyle\sum_{e=1}^{n^{2}-1}d_{ade}d_{bce}
\end{equation}
\section{Examples}
Now, we have some theorems and relations on the generalized
Gell-Mann matrices which we need for expressing a tensor permutation
matrix in terms of the generalized Gell-Mann matrices. In this
section, we treat some examples.
\subsection{$U_{n^{\otimes3}}(\sigma)$}

1) $\sigma=(1\;2\;3)$\\
By employing the Lemma \ref{thm33}\\
\begin{center}
 $\sigma=(1\;2)(2\;3)$
\end{center}
and by using the Theorem \ref{thm34}
\begin{equation}\label{e51}
U_{n^{\otimes3}}\left((1\;2\;3)\right)=U_{n^{\otimes3}}\left((1\;2)\right)\cdot
U_{n^{\otimes3}}\left((2\;3)\right)
\end{equation}
 However, \cite{rak05}
\begin{equation*}
U_{n\otimes n}\;=\;\frac{1}{n}I_{n}\otimes
I_{n}+\frac{1}{2}\displaystyle\sum_{a=1}^{n^{2}-1}\Lambda_{a}\otimes\Lambda_{a}
\end{equation*}
Then,
\begin{equation*}
 U_{n^{\otimes3}}\left((1\;2)\right)=\;\frac{1}{n}I_{n}\otimes
I_{n}\otimes
I_{n}+\frac{1}{2}\displaystyle\sum_{a=1}^{n^{2}-1}\Lambda_{a}\otimes\Lambda_{a}\otimes
I_{n}
\end{equation*}
and
\begin{equation*}
 U_{n^{\otimes3}}\left((2\;3)\right)=\;\frac{1}{n}I_{n}\otimes
I_{n}\otimes
I_{n}+\frac{1}{2}\displaystyle\sum_{a=1}^{n^{2}-1}I_{n}\otimes\Lambda_{a}\otimes\Lambda_{a}
\end{equation*}
So, by using (\ref{e51}) we have
\begin{multline*}
U_{n^{\otimes3}}\left((1\;2\;3)\right)=\frac{1}{n^{2}}I_{n}\otimes
I_{n}\otimes
I_{n}+\frac{1}{2n}\displaystyle\sum_{a=1}^{n^{2}-1}I_{n}\otimes\Lambda_{a}\otimes\Lambda_{a}\\
+\frac{1}{2n}\displaystyle\sum_{a=1}^{n^{2}-1}\Lambda_{a}\otimes\Lambda_{a}\otimes
I_{n}+\frac{1}{4}\displaystyle\sum_{a=1}^{n^{2}-1}\displaystyle\sum_{b=1}^{n^{2}-1}
\Lambda_{a}\otimes\Lambda_{a}\Lambda_{b}\otimes\Lambda_{b}
\end{multline*}
Hence, employing the relation (\ref{e43})
\begin{multline}\label{e52}
U_{n^{\otimes3}}\left((1\;2\;3)\right)=\frac{1}{n^{2}}I_{n}\otimes
I_{n}\otimes I_{n}+\frac{1}{2n}\displaystyle\sum_{a=1}^{n^{2}-1}I_{n}\otimes\Lambda_{a}\otimes\Lambda_{a}\\
+\frac{1}{2n}\displaystyle\sum_{a=1}^{n^{2}-1}\Lambda_{a}\otimes\Lambda_{a}\otimes
I_{n}+\frac{1}{2n}\displaystyle\sum_{a=1}^{n^{2}-1}\Lambda_{a}\otimes
I_{n}\otimes\Lambda_{a}\\
-\frac{i}{4}\displaystyle\sum_{a=1}^{n^{2}-1}\displaystyle\sum_{b=1}^{n^{2}-1}
\displaystyle\sum_{c=1}^{n^{2}-1}f_{abc}\Lambda_{a}\otimes\Lambda_{b}\otimes\Lambda_{c}
+\frac{1}{4}\displaystyle\sum_{a=1}^{n^{2}-1}\displaystyle\sum_{b=1}^{n^{2}-1}
\displaystyle\sum_{c=1}^{n^{2}-1}d_{abc}\Lambda_{a}\otimes\Lambda_{b}\otimes\Lambda_{c}
\end{multline}
2) $\sigma=(1\;3\;2)$\\
By employing the Lemma \ref{thm33},\\
\begin{center}
$\sigma=(1\;3)(3\;2)$\\
\end{center}
 and by using the Theorem \ref{thm21}, we have
\begin{equation*}
U_{n^{\otimes3}}\left((1\;3)\right)=\frac{1}{n}I_{n}\otimes
I_{n}\otimes
I_{n}+\frac{1}{2}\displaystyle\sum_{a=1}^{n^{2}-1}\Lambda_{a}\otimes
I_{n}\otimes\Lambda_{a}
\end{equation*}
and by using the same method
\begin{multline*}
U_{n^{\otimes3}}\left((1\;3\;2)\right)=\frac{1}{n^{2}}I_{n}\otimes
I_{n}\otimes I_{n}+\frac{1}{2n}\displaystyle\sum_{a=1}^{n^{2}-1}I_{n}\otimes\Lambda_{a}\otimes\Lambda_{a}\\
+\frac{1}{2n}\displaystyle\sum_{a=1}^{n^{2}-1}\Lambda_{a}\otimes\Lambda_{a}\otimes
I_{n}+\frac{1}{2n}\displaystyle\sum_{a=1}^{n^{2}-1}\Lambda_{a}\otimes
I_{n}\otimes\Lambda_{a}\\
+\frac{i}{4}\displaystyle\sum_{a=1}^{n^{2}-1}\displaystyle\sum_{b=1}^{n^{2}-1}
\displaystyle\sum_{c=1}^{n^{2}-1}f_{abc}\Lambda_{a}\otimes\Lambda_{b}\otimes\Lambda_{c}
+\frac{1}{4}\displaystyle\sum_{a=1}^{n^{2}-1}\displaystyle\sum_{b=1}^{n^{2}-1}
\displaystyle\sum_{c=1}^{n^{2}-1}d_{abc}\Lambda_{a}\otimes\Lambda_{b}\otimes\Lambda_{c}
\end{multline*}
The difference between $U_{n^{\otimes3}}\left((1\;2\;3)\right)$ and
$U_{n^{\otimes3}}\left((1\;3\;2)\right)$ is the minus sign before
the fifth term.
\subsection{$U_{n^{\otimes4}}(\sigma)$, $\sigma=(1\;2\;3\;4)$}
By employing the Lemma \ref{thm33},\\
\begin{center}
$\sigma=(1\;2\;3)(3\;4)$\\
\end{center}
and by using the formula (\ref{e52}), Theorem \ref{thm34} and the
proposition \ref{thm11}  , we have
\begin{equation*}
\begin{split}
U_{n^{\otimes4}}(\sigma)&=\frac{1}{n^{3}}I_{n}\otimes I_{n}\otimes
I_{n}\otimes I_{n}
+\frac{1}{2n^{2}}\displaystyle\sum_{a=1}^{n^{2}-1}I_{n}\otimes\Lambda_{a}\otimes\Lambda_{a}\otimes
I_{n}
+\frac{1}{2n^{2}}\displaystyle\sum_{a=1}^{n^{2}-1}\Lambda_{a}\otimes\Lambda_{a}\otimes
I_{n}\otimes I_{n}\\
&+\frac{1}{2n^{2}}\displaystyle\sum_{a=1}^{n^{2}-1}\Lambda_{a}\otimes
I_{n}\otimes\Lambda_{a}\otimes
I_{n}+\frac{1}{2n^{2}}\displaystyle\sum_{a=1}^{n^{2}-1}I_{n}\otimes\ I_{n}\otimes\Lambda_{a}\otimes\Lambda_{a}\\
&+\frac{1}{4n}\displaystyle\sum_{a=1}^{n^{2}-1}\displaystyle\sum_{b=1}^{n^{2}-1}
\Lambda_{a}\otimes\Lambda_{a}\otimes\Lambda_{b}\otimes\Lambda_{b}
+\frac{1}{4n}\displaystyle\sum_{a=1}^{n^{2}-1}\displaystyle\sum_{b=1}^{n^{2}-1}
I_{n}\otimes\Lambda_{a}\otimes\Lambda_{a}\Lambda_{b}\otimes\Lambda_{b}\\
&+\frac{1}{4n}\displaystyle\sum_{a=1}^{n^{2}-1}\displaystyle\sum_{b=1}^{n^{2}-1}
\Lambda_{a}\otimes
I_{n}\otimes\Lambda_{a}\Lambda_{b}\otimes\Lambda_{b}\\
&-\frac{i}{4n}\displaystyle\sum_{a=1}^{n^{2}-1}\displaystyle\sum_{b=1}^{n^{2}-1}\displaystyle\sum_{c=1}^{n^{2}-1}
f_{abc}\Lambda_{a}\otimes\Lambda_{b}\otimes\Lambda_{c}\otimes
I_{n}\\
&-\frac{i}{8}\displaystyle\sum_{a=1}^{n^{2}-1}\displaystyle\sum_{b=1}^{n^{2}-1}\displaystyle\sum_{c=1}^{n^{2}-1}
\displaystyle\sum_{e=1}^{n^{2}-1}f_{abc}\Lambda_{a}\otimes\Lambda_{b}\otimes\Lambda_{c}\Lambda_{e}\otimes\Lambda_{e}\\
&+\frac{1}{4n}\displaystyle\sum_{a=1}^{n^{2}-1}\displaystyle\sum_{b=1}^{n^{2}-1}\displaystyle\sum_{c=1}^{n^{2}-1}
d_{abc}\Lambda_{a}\otimes\Lambda_{b}\otimes\Lambda_{c}\otimes
I_{n}\\
&+\frac{1}{8}\displaystyle\sum_{a=1}^{n^{2}-1}\displaystyle\sum_{b=1}^{n^{2}-1}\displaystyle\sum_{c=1}^{n^{2}-1}
\displaystyle\sum_{e=1}^{n^{2}-1}d_{abc}\Lambda_{a}\otimes\Lambda_{b}\otimes\Lambda_{c}\Lambda_{e}\otimes\Lambda_{e}
\end{split}
\end{equation*}
By using the relation (\ref{e43}) and the relation (\ref{e44}), we
have
\begin{equation*}
\begin{split}
U_{n^{\otimes4}}(\sigma)&=\frac{1}{n^{3}}I_{n}\otimes I_{n}\otimes
I_{n}\otimes I_{n}
+\frac{1}{2n^{2}}\displaystyle\sum_{a=1}^{n^{2}-1}I_{n}\otimes\Lambda_{a}\otimes\Lambda_{a}\otimes
I_{n}
+\frac{1}{2n^{2}}\displaystyle\sum_{a=1}^{n^{2}-1}\Lambda_{a}\otimes\Lambda_{a}\otimes
I_{n}\otimes I_{n}\\
&+\frac{1}{2n^{2}}\displaystyle\sum_{a=1}^{n^{2}-1}\Lambda_{a}\otimes
I_{n}\otimes\Lambda_{a}\otimes
I_{n}+\frac{1}{2n^{2}}\displaystyle\sum_{a=1}^{n^{2}-1}I_{n}\otimes\ I_{n}\otimes\Lambda_{a}\otimes\Lambda_{a}\\
&+\frac{1}{2n^{2}}\displaystyle\sum_{a=1}^{n^{2}-1}I_{n}\otimes\Lambda_{a}\otimes
I_{n}\otimes\Lambda_{a}+\frac{1}{2n^{2}}\displaystyle\sum_{a=1}^{n^{2}-1}\Lambda_{a}\otimes
I_{n}\otimes I_{n}\otimes\Lambda_{a}\\
&+\frac{1}{4n}\displaystyle\sum_{a=1}^{n^{2}-1}\displaystyle\sum_{b=1}^{n^{2}-1}
\Lambda_{a}\otimes\Lambda_{a}\otimes\Lambda_{b}\otimes\Lambda_{b}+\frac{1}{4n}\displaystyle\sum_{a=1}^{n^{2}-1}\displaystyle\sum_{b=1}^{n^{2}-1}
\Lambda_{a}\otimes\Lambda_{b}\otimes\Lambda_{b}\otimes\Lambda_{a}\\
&-\frac{1}{4n}\displaystyle\sum_{a=1}^{n^{2}-1}\displaystyle\sum_{b=1}^{n^{2}-1}
\Lambda_{a}\otimes\Lambda_{b}\otimes\Lambda_{a}\otimes\Lambda_{b}\\
&+\frac{1}{4n}\displaystyle\sum_{a=1}^{n^{2}-1}\displaystyle\sum_{b=1}^{n^{2}-1}
\displaystyle\sum_{c=1}^{n^{2}-1}d_{abc}I_{n}\otimes\Lambda_{a}\otimes\Lambda_{b}\otimes\Lambda_{c}
-\frac{i}{4n}\displaystyle\sum_{a=1}^{n^{2}-1}\displaystyle\sum_{b=1}^{n^{2}-1}
\displaystyle\sum_{c=1}^{n^{2}-1}f_{abc}I_{n}\otimes\Lambda_{a}\otimes\Lambda_{b}\otimes\Lambda_{c}\\
&+\frac{1}{4n}\displaystyle\sum_{a=1}^{n^{2}-1}\displaystyle\sum_{b=1}^{n^{2}-1}
\displaystyle\sum_{c=1}^{n^{2}-1}d_{abc}\Lambda_{a}\otimes
I_{n}\otimes\Lambda_{b}\otimes\Lambda_{c}
-\frac{i}{4n}\displaystyle\sum_{a=1}^{n^{2}-1}\displaystyle\sum_{b=1}^{n^{2}-1}
\displaystyle\sum_{c=1}^{n^{2}-1}f_{abc}\Lambda_{a}\otimes I_{n}\otimes\Lambda_{b}\otimes\Lambda_{c}\\
&+\frac{1}{4n}\displaystyle\sum_{a=1}^{n^{2}-1}\displaystyle\sum_{b=1}^{n^{2}-1}
\displaystyle\sum_{c=1}^{n^{2}-1}d_{abc}\Lambda_{a}\otimes\Lambda_{b}\otimes\Lambda_{c}\otimes
I_{n}
-\frac{i}{4n}\displaystyle\sum_{a=1}^{n^{2}-1}\displaystyle\sum_{b=1}^{n^{2}-1}
\displaystyle\sum_{c=1}^{n^{2}-1}f_{abc}\Lambda_{a}\otimes\Lambda_{b}\otimes\Lambda_{c}\otimes
I_{n}\\
&+\frac{1}{4n}\displaystyle\sum_{a=1}^{n^{2}-1}\displaystyle\sum_{b=1}^{n^{2}-1}
\displaystyle\sum_{c=1}^{n^{2}-1}d_{abc}\Lambda_{a}\otimes\Lambda_{b}\otimes
I_{n}\otimes\Lambda_{c}
-\frac{i}{4n}\displaystyle\sum_{a=1}^{n^{2}-1}\displaystyle\sum_{b=1}^{n^{2}-1}
\displaystyle\sum_{c=1}^{n^{2}-1}f_{abc}\Lambda_{a}\otimes\Lambda_{b}\otimes
I_{n}\otimes\Lambda_{c}\\
&+\frac{1}{8}\displaystyle\sum_{a=1}^{n^{2}-1}\displaystyle\sum_{b=1}^{n^{2}-1}\displaystyle\sum_{c=1}^{n^{2}-1}
\displaystyle\sum_{e=1}^{n^{2}-1}\displaystyle\sum_{g=1}^{n^{2}-1}\left(-if_{abc}d_{ceg}+id_{abc}f_{ceg}+d_{aec}d_{bgc}
-d_{agc}d_{cbe}+d_{abc}d_{ceg}\right)\\
&\Lambda_{a}\otimes\Lambda_{b}\otimes\Lambda_{g}\otimes\Lambda_{e}\\
\end{split}
\end{equation*}
\subsection{$U_{2^{\otimes3}}(\sigma)$, $\sigma\in S_{3}$ }
Now we give the formulae giving $U_{2^{\otimes3}}(\sigma)$,
naturally in terms of the Pauli matrices
\begin{center}$\sigma_1$ = $\left(
\begin{array}{cc}
  0 & 1 \\
  1 & 0 \\
\end{array}
\right)$, $\sigma_2$\;=\;$\left(
\begin{array}{cc}
  0 & -i \\
  i & 0 \\
\end{array}
\right)$, $\sigma_3$\;=\;$\left(
\begin{array}{cc}
  1 & 0 \\
  0 & -1 \\
\end{array}
\right)$
\end{center}
 Using the relation \cite{rao72}
\begin{equation*}
\sigma_{l}\sigma_{k}=\delta_{lk}I_{2}+i\displaystyle\sum_{m=1}^{3}\varepsilon_{lkm}\sigma_{m}
\end{equation*}
where $\varepsilon_{ijk}$ is totally antisymmetric in the three
indices, which is equal 1 if $(i\;\;j\;\;k)=(1\;\;2\;\;3)$, we have
\begin{equation*}
\begin{split}
U_{2^{\otimes3}}(1\;\;2\;\;3)&=\frac{1}{4}I_{2}\otimes I_{2}\otimes
I_{2}
+\frac{1}{4}\displaystyle\sum_{l=1}^{3}I_{2}\otimes\sigma_{l}\otimes\sigma_{l}
+\frac{1}{4}\displaystyle\sum_{l=1}^{3}\sigma_{l}\otimes
I_{2}\otimes\sigma_{l}\\
&+\frac{1}{4}\displaystyle\sum_{l=1}^{3}\sigma_{l}\otimes\sigma_{l}\otimes
I_{2}-\frac{i}{4}\displaystyle\sum_{i=1}^{3}\displaystyle\sum_{j=1}^{3}\displaystyle\sum_{k=1}^{3}
\varepsilon_{ijk}\sigma_{i}\otimes\sigma_{j}\otimes\sigma_{k}
\end{split}
\end{equation*}
and
\begin{equation*}
\begin{split}
U_{2^{\otimes3}}(1\;\;3\;\;2)&=\frac{1}{4}I_{2}\otimes I_{2}\otimes
I_{2}
+\frac{1}{4}\displaystyle\sum_{l=1}^{3}I_{2}\otimes\sigma_{l}\otimes\sigma_{l}
+\frac{1}{4}\displaystyle\sum_{l=1}^{3}\sigma_{l}\otimes
I_{2}\otimes\sigma_{l}\\
&+\frac{1}{4}\displaystyle\sum_{l=1}^{3}\sigma_{l}\otimes\sigma_{l}\otimes
I_{2}+\frac{i}{4}\displaystyle\sum_{i=1}^{3}\displaystyle\sum_{j=1}^{3}\displaystyle\sum_{k=1}^{3}
\varepsilon_{ijk}\sigma_{i}\otimes\sigma_{j}\otimes\sigma_{k}
\end{split}
\end{equation*}
\section*{Conclusion}
Based on the fact that a tensor permutation matrix is a product of
tensor transposition matrices and on the Theorem \ref{thm21}, with
the help of the expression of a tensor commutation matrix in terms
of the generalized Gell-Mann matrices, we can express a tensor
permutation matrix as linear combination of the tensor products of
the generalized Gell-Mann matrices.\\
\indent We have no intention for searching a general formula.
However, we have shown that any tensor permutation matrix can be
expressed in terms of the generalized Gell-Mann matrices and then
the expression can be simplified by using the relations between
these Matrices.
\subsection*{Acknowledgements} The author would like to thank
Ratsimbarison Mahasedra for encouragement and for critical reading
the manuscript.

\renewcommand{\bibname}{References}

\end{document}